\documentclass{appolb}
\usepackage{epsfig}
\usepackage{amsmath,amssymb}

\newcommand{\beqn}{\begin{equation}}
\newcommand{\eeqn}{\end{equation}}

\begin{document}
\title{\vspace*{-2cm} Planetary Impacts by Clustered Quark Matter Strangelets}
\author{Lance Labun and Jan Rafelski
\address{Department of Physics, The University of Arizona, Tucson, 85721 USA}
}
\date{20 December 2011}
\maketitle

\section{Overview}
Large mass strangelets $10^{-12}<M/M_{\oplus}<10^{-5}$ ($M_{\oplus}=6\:10^{24}~{\rm kg}$ is the Earth's mass) could originate either in the early universe phase transition to the hadronic vacuum~\cite{EUstrangelets} or in collisions of stellar strange quark matter objects~\cite{fragstrangelets}. Isolated $s\simeq u\simeq d$-quark matter  strangelets could be stable for a large range of baryon numbers~\cite{Madsen99}.  

Despite several mechanisms by which the strange quark matter can evaporate, strangelets have been argued to survive in the present time~\cite{Madsen99,evapstrangelets}.  Stability of bulk strange quark matter is made possible by the reduction in free energy per baryon due to the presence of the third Fermi sea.  At sufficiently high density, this reduction may lead to a new minimum, making strange quark matter energetically favored.  Here we show a new minimum occurs in the case of clustered quark matter~\cite{Clark:1986cq}, opening the opportunity to discover large strangelets' interactions with bodies in the solar system. 

Impacts by strangelets with meteorite-scale mass and smaller $M<10^{-12}M_{\oplus}$ generate seismic disturbances on passing through a planet~\cite{strangeletquakes}.  Seismic events fitting the profile of strangelet passage have been sought on the Earth~\cite{Herrin:2005kb} and Moon~\cite{Banerdt:2006}.  While seismic searches must record the strangelet passage event in real time, persistence over geologic timescales is possible for impact signatures of larger strange\-lets with mass in the asteroid mass domain $10^{-12}<M/M_{\oplus}<0.01$. Such large strangelets have gravitational tidal interactions with normal matter, capable of extensively disrupting the interior and surface of rocky solar bodies and leading to long-lived impact features.

\section{Clustered strange quark matter}
A smaller strange quark mass improves stability of strangelets by reducing the energy penalty resulting from the mass difference.  Interest in strange quark matter and strangelets is thus heightened by the reduction in the official strange quark mass range $m_s(2\:{\rm GeV})=104^{+26}_{-34}~\text{MeV}$~\cite{PDG2010} due to the most recent lattice results~\cite{mslattice}.  With this value of $m_s$, calculations within the framework of QCD~\cite{pQCDSQM} lead to the expectation of strange quark matter in the center of neutron stars, given observations of neutron stars with masses greater than two times the solar mass.

To model non-perturbative effects of QCD at low temperature and near nuclear density, we consider the clustered quark model~\cite{Clark:1986cq} (CQM) extended to include a massive strange quark.  In CQM the strong interactions are modeled with correlated 3-quark colorless states in a sea of free quarks.  By taking 3-quark states with the quantum numbers and qualitative spectrum properties of nucleons in the confining (hadronic) phase, the model phenomenologically interpolates between the hadronic phase and the quark matter phase.

We extend the CQM incorporating the massive strange quark by including the third Fermi sea of free massive strange quarks as well as correlated 3-quark states corresponding to the four lowest mass strange $(S=-1)$ baryons, the isospin singlet $\Lambda$ and isospin triplet $\Sigma^{\pm,0}$.   The strangeness content is controlled by chemical equilibrium between quark flavors, which is enforced by weak interactions, and the state of the system is further constrained by the baryon number.  The thermodynamic potential can then be computed in the excluded volume method as a function of baryon density.

\begin{figure}
\centerline{
\includegraphics[width=0.7\textwidth]{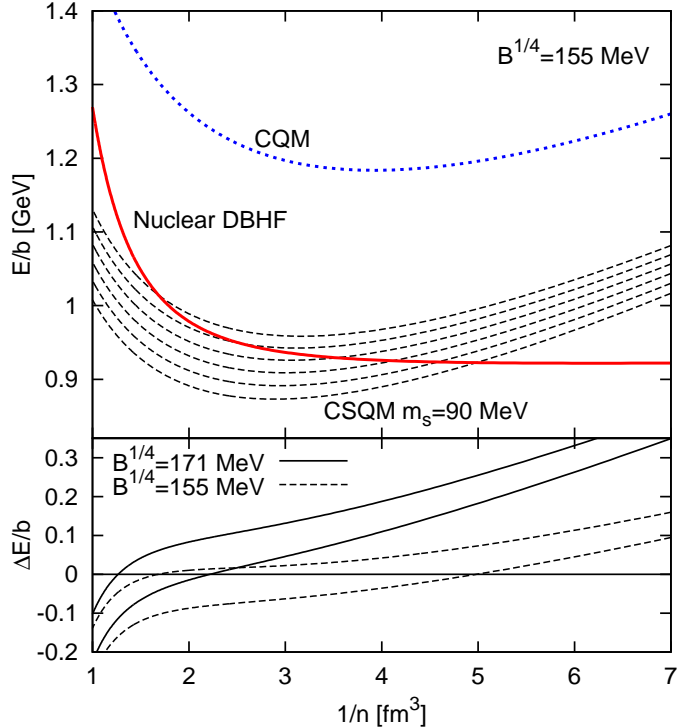}}
\caption{Top frame: Energy per baryon $E/b$ as a function of baryon density $n$.  The lowest dashed curve is for clustered strange quark matter (CSQM) with $m_s=90~{\rm MeV}$ and each successive curve increases $m_s$ by 10 MeV, ending with $m_s=140~{\rm MeV}$ for the uppermost dashed curve. Comparison curves for $u,d$-only clustered quark matter and Dirac-Brueckner-Hartree-Fock (DBHF)~\cite{nucmatter}  are also shown. Bottom frame: the energy difference between CSQM and  DBHF, with upper curve corresponding to $m_s=140~{\rm MeV}$ and lower corresponding to $m_s=90~{\rm MeV}$.  $B^{1/4}$=155 MeV for dashed lines and =171 MeV for solid lines.  \label{fig:clusteredsqm}}
\end{figure}

The initial and still preliminary results from this study suggest relatively strong binding due to the added strange flavor.  Figure~\ref{fig:clusteredsqm} shows the energy per baryon of bulk clustered strange quark matter (CSQM) $E/b=3P+B$ with bag constant $B^{1/4}=155~{\rm MeV}$ and constant strange quark masses increasing from $m_s=90~{\rm MeV}$.  The strange quark mass varies with the energy scale (here set by chemical potential), with $m_s\simeq 140$ MeV  below nuclear density and decreasing for higher densities.  The semi-realistic value $B^{1/4}=155~{\rm MeV}$, corresponding to $4B=0.247$ GeV/fm$^3$, has been selected to ensure that normal nuclear matter near $1/n=6\:{\rm fm}^3$  described within Dirac-Brueckner-Hartree-Fock (DBHF) model~\cite{nucmatter} is sufficiently stable against decay into SQM for physical values of $m_s$. As bottom frame shows, an increase to $B^{1/4}=171~{\rm MeV}$ reduces CSQM binding, but still results in an absolutely bound state.

To assess the density at which transition to CSQM could occur we show in the lower frame of Figure~\ref{fig:clusteredsqm} the difference
$\Delta E/b= 3P_{\rm CSQM}+B-(E/b)_{\rm DBHF}$
i.e. the energy per baryon of CSQM compared to DBHF nuclear matter~\cite{nucmatter}. Clustered strange quark matter appears energetically favored in bulk at zero temperature, $\Delta E/b$ being negative for densities  above nuclear density $1/n<5\:{\rm fm}^3$. Comparison of $B^{1/4}=155~{\rm MeV}$ (dashed lines) and $B^{1/4}=171~{\rm MeV}$ (solid lines) shows the outcome is sensitive to the value of the bag constant $B$.  Our results suggest that an absolutely stable minimum in $E/b$ always exists within CSQM, which means that such strangelets once produced never decay.

For very large strangelets, stability is enhanced by gravity, while for (much) smaller strange\-lets, stability depends also on the surface tension between the quark matter phase and external hadronic vacuum~\cite{SQMsurftension}.

\section{Impacts on Planetary Bodies}
A strangelet is a prominent representative of a compact ultra dense object (CUDO).  Features of CUDO impacts on rocky solar system bodies were recently explored for a CUDO mass range $10^{-12}<M/M_{\oplus}<0.01$ in~\cite{Labun11}.  The compactness of the impactor results in a steep gradient in the gravitational potential, and the gravitational tidal force of the strangelet is strong enough to pulverize normal matter.  For these reasons, a beyond-nuclear density impactor cannot be stopped at the surface of the planet, and rather passes through the planet, creating both entry and exit features on the surface.

The strong gravitational tidal forces result in two primary mechanisms for depositing the energy of the impacting strangelet in the planet.  The first is heating, entrainment, and mixing of rock in the planet's interior due to pulverization, accretion, and displacement of the rock nearest the strangelet trajectory.  The second is the seismic shock wave created by tidal stresses farther from the strangelet.  Where this shock wave meets the surface, the topography will be extensively disturbed, creating further debris that can be pulled into the atmosphere.

Impacts by intermediate mass strangelets could resolve a variety of geologic puzzles both on the Earth and on other solar system bodies~\cite{Labun11}:\\
1. The entrainment of rock during the strangles passage through the interior leads to mixing of material across all layers of the impacted body.  Non-volatile elements, differentiated into the core during the Earth's molten early stages, could thus be reintroduced into the upper mantle and crust~\cite{mixing} where we find these today.\\
2. The exit of the strangelet could seed mantle plumes~\cite{plumes} and establish punctual volcanic hotspots in the middle of Earth's thick continental crust.  Moreover, when the strangelet passes near to the planetary core, the hotspots are seeded by magma of a deep origin.  Basing plumes and hotspots deep in the mantle is difficult in the present theory of mantle dynamics, yet is determined to so in the case of the Hawaiian hotspot~\cite{Hawaii}.\\
3. The ability of strangelets to pull material from the surface (high) into the atmosphere offers an effective mechanism to create ``nuclear winter'' conditions.  For example, a global dust veil occurring in AD 536 is not satisfactorily explained as either a comet impact~\cite{AD536comet} or a volcanic eruption~\cite{AD536volcano}.\\
4. Penetration of the mantle and deposition of debris high in the atmosphere on exit would provide a causal link between impact signals, large volume volcanic eruptions and mass extinctions~\cite{extinctions}.  Large, explosive volcanic events~\cite{supereruptions}, also capable of ejecting matter high into the atmosphere, are believed to occur at a frequency similar to the expected CUDO impact rate, and could potentially be just that.\\
5. Causing lava flow on planetary bodies considered volcanically inactive, CUDO impacts could contribute to the geologically recent volcanic formations observed on the Moon~\cite{Moonvolcanoes}, Mercury~\cite{MESSENGER} and Mars~\cite{Marsvolcanoes}.  Only the presence of an ultra-dense component (such as a strangelet) in the impactor can cause lava flow; normal matter meteorites are incapable of penetrating the crust~\cite{Ivanov:2003}. 

Naturally, impacts by larger mass strangelets are expected to occur more rarely than impacts by smaller mass strangelets, which are recognized by acoustic signatures.  Preservation of impact features over geological time scale may sufficiently enhance detection sensitivity.  However, even if agreement should be reached that a CUDO  impact has occurred, features specific to strangelets will need to be recognized. 

\section{Summary and Conclusions}
We have discussed the increased  stability of strange quark matter given  improved understanding of strange quark mass and clustering feature of cold quark matter. Our result implies that nuclear matter could be converted to strange quark matter by compression, e.g. in relativistic heavy ion  collisions, or more likely, in the center of neutron stars.

Many laboratory searches for small strangelets have produced no positive signature to date~\cite{Sandweiss:2004bu}. We hope that  persistence of impact formations over geologic timescales which turn rocky planets into integrating detectors will  open the search for large strangelets. \\[3mm]

\noindent This work was supported by  the grant from the U.S. Department of Energy, DE-FG02-04ER41318.


\end{document}